\documentclass[a4paper,11pt]{article}
\usepackage{pos}

\title{Competition between diffusion and rapid expansion and its impact on critical fluctuations in heavy-ion collisions}
\ShortTitle{Impact of diffusion versus rapid expansion on critical fluctuations}

\author*[a]{Grégoire Pihan}
\author[a]{Marcus Bluhm}
\author[b,c]{Masakiyo Kitazawa}
\author[a]{Taklit Sami}
\author[a]{Marlene Nahrgang}
\affiliation[a]{SUBATECH UMR 6457 (IMT Atlantique, Université de Nantes, IN2P3/CNRS),\\
  4 rue Alfred Kastler 44307, Nantes, France}
\affiliation[b]{Department of Physics, Osaka University,\\
  Toyonaka, Osaka 560-0043, Japan}
\affiliation[c]{J-PARC Branch, KEK Theory Center, Institute of Particle and Nuclear Studies, \\ KEK, 203-1, Shirakata, Tokai, Ibaraki, 319-1106, Japan}

\emailAdd{pihan@subatech.in2p3.fr}
\emailAdd{bluhm@subatech.in2p3.fr}
\emailAdd{kitazawa@phys.sci.osaka-u.ac.jp}
\emailAdd{sami@subatech.in2p3.fr}
\emailAdd{nahrgang@subatech.in2p3.fr}

\abstract{We study the impact of the competition between the expansion of the medium created in heavy-ion collisions and its diffusive properties on the critical fluctuations of the net-baryon density. As the relaxation time of the fluctuations is connected with the diffusive properties, the latter determine the in- or out-of-equilibrium nature of the net-baryon density fluctuations during the fireball evolution. This may result in important consequences for the phenomenological interpretation of heavy-ion collision data. We study three possible situations that can occur as a result of the competition between diffusion and expansion and discuss the impact of this competition on the integrated second and fourth order cumulants of net-baryon density fluctuations at freeze-out in these situations.}

\FullConference{%
	The International conference on Critical Point and Onset of Deconfinement - CPOD2021\\
  	15 – 19 March 2021\\
 	Online - zoom
 }

\begin{document}
\maketitle

\section{Introduction}
\label{sec:I}

Equilibrium fluctuations of the conserved charges in QCD, such as the net-baryon number, have been proposed as crucial observables to trace the phase transition and to probe criticality in the QCD phase diagram~\cite{Asakawa:2015ybt}. However, the highly dynamical context of a heavy-ion collision prevents a direct comparison between equilibrium predictions and experimental results as the violent expansion of the system may drive the fluctuations out of equilibrium~\cite{Bluhm:2020mpc}. In turn, a relevant dynamical study of the fluctuations to evaluate their distance from equilibrium is needed to interpret experimental measurements. Due to the connection between the relaxation time and the diffusive properties of the medium, this distance is determined by the competition between the diffusion and the expansion. In this work we distinguish three regimes: First, when the diffusion wins over the expansion the fluctuations are always close to equilibrium and the critical enhancement is large but short-lived. Second, when the expansion wins over the diffusion, the critical signal is washed out by the expansion. Third, an intermediate state with smaller criticality but a longer survival of the signal in the fluctuations. To study these three regimes and to discuss the impact on the phenomenology of heavy-ion collisions we make use of a stochastic diffusion equation (SDE) expressed in Milne coordinates for a Bjorken-type expansion of the medium~\cite{Kitazawa:2020kvc}. The potential in the SDE is derived from second and fourth order susceptibilities per unit of rapidity. The latter allows us to connect the criticality from the 3D Ising model in the scaling region with lattice QCD calculations far from it.

\section{Diffusive dynamics of net-baryon density fluctuations}
\label{sec:II}

\begin{figure}[t]
    \centering
    \includegraphics[width=0.45\textwidth,clip]{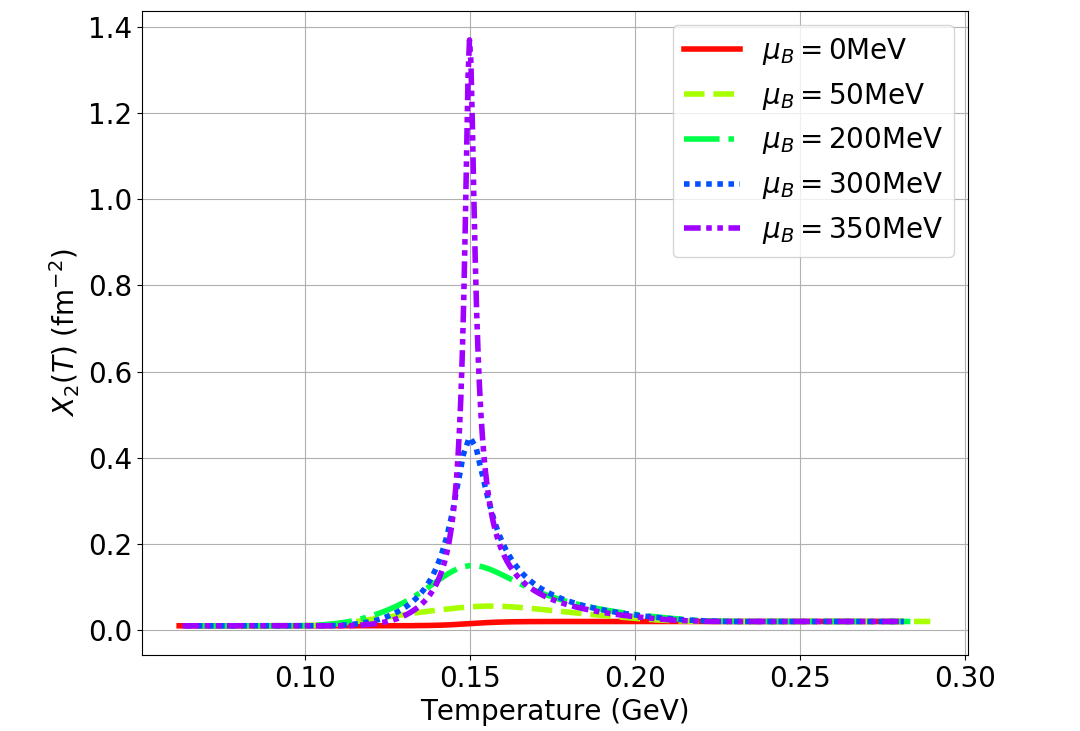}
    \hspace{0.03\textwidth}
    \includegraphics[width=0.45\textwidth,clip]{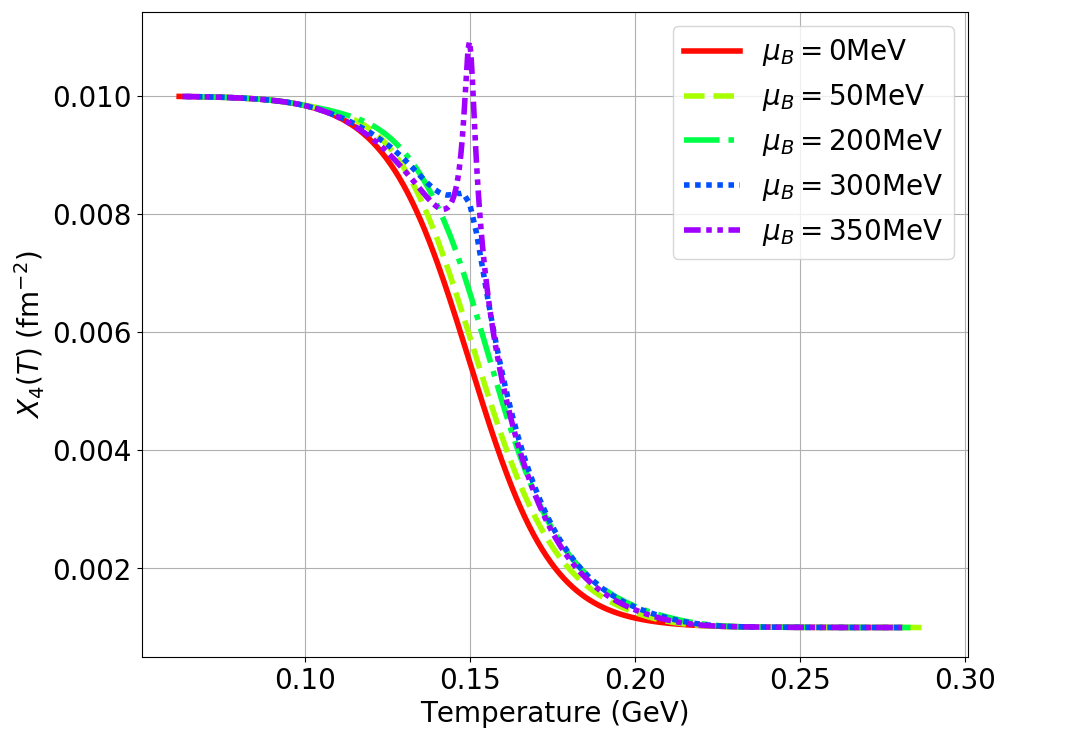}
    \caption{The second (left panel) and fourth order (right panel) susceptibilities per unit of rapidity as a function of temperature $T$ for different baryon-chemical potentials $\mu_B$.}
    \label{fig:param}
\end{figure}

The diffusive dynamics of net-baryon density fluctuations in a rapidly expanding medium can be studied via a non-linear, stochastic diffusion equation expressed in the Milne coordinates $\tau$ and~$y$ reading 
 
\begin{equation}
	\partial_{\tau} \tilde{n}_B = D(\tau) \,\partial_y^2 \tilde{n}_B  
        - \frac{D K}{n_c \tau^3} \partial_y^4 \tilde{n}_B+ \frac{D \lambda_4(\tau)}{n_c \tau} \partial_y^2 {\tilde{n}_B}^3 - \partial_y \xi,
\label{eq:FullSDE}
\end{equation}
where $D(\tau) = D m^2(\tau)/n_c$, $\tilde{n}_B(\tau,y) = (n_B(\tau,y) - n_c)/n_c$ is the field of reduced net-baryon density fluctuations around a critical density $n_c$, $D$ is the diffusion length and $K$ the surface tension. The white noise field $\xi(\tau,y)$ respects the fluctuation-dissipation balance locally following 
\begin{equation}
        \langle \xi(\tau_1, y_1) \xi(\tau_2, y_2) \rangle = \frac{2 D }{n_c \tau} \delta(\tau_1 - \tau_2) \delta(y_1 - y_2) \,.
\label{eq:FlucDissBalance}
\end{equation}
The effective mass $m^2$ entering the definition of $D(\tau)$ and the non-linear coupling $\lambda_4$ are related to the second and fourth order susceptibilities per unit of rapidity for given $T$ as function of $\tau$ via 
\begin{equation}
	m^2(T) = \frac{n_c^2}{\chi_2(T)} \,, \quad 
	\lambda_4(T) = \frac{n_c^6}{6 \chi_4(T)}.
\label{eq:chis}
\end{equation}
In Fig.~\ref{fig:param} we show the temperature dependence of the second and fourth order susceptibilities per unit of rapidity, $\chi_2(T)$ and $\chi_4(T)$, along different trajectories of constant $\mu_B$. The parametrizations are obtained by adding singular contributions, which stem from a matching to the susceptibilities of the 3D Ising model scaling equation of state, and regular contributions similar to the procedure proposed in~\cite{Sakaida:2017rtj} and in line with results from lattice QCD, see~\cite{PihanFuture} for more details. 

\section{Diffusion versus expansion}
\label{sec:III}

\begin{figure}[t!]
    \centering
    \includegraphics[scale=0.24]{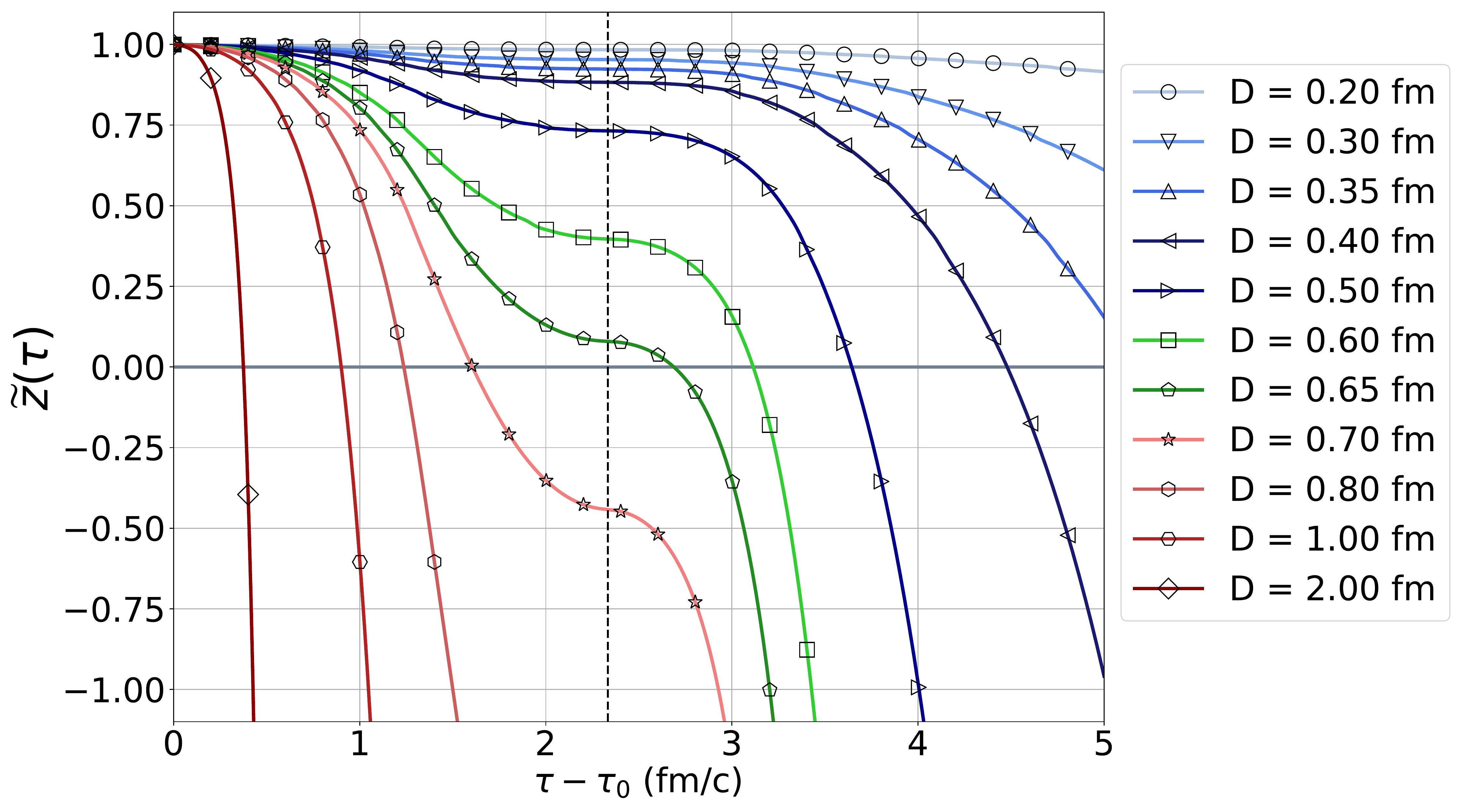}
        \caption{The normalized 
        difference between the size of the expanding system $z_{max}(\tau)$ and the time evolution of the standard deviation of an initially Gaussian density profile in Cartesian coordinates $z_D(\tau)$ as a function of proper time and input diffusion length $D$ for the deterministic evolution of the net-baryon density. The vertical dashed line corresponds to the critical proper time $\tau_c$.}
    \label{fig:deterministicD}
\end{figure}

The setup presented in the previous section reproduces the dynamical scaling of model B for non-expanding systems~\cite{Nahrgang:2020yxm}. Accordingly, the relaxation time and the diffusion term (first term on the R.H.S. of Eq.~\eqref{eq:FullSDE}) are related, $\tau_{\rm relax} \propto D(\tau)^{-1}$. Thus, the ability of the net-baryon density fluctuations to reach equilibrium or not during the expansion can be studied by looking at the competition between the diffusion and the expansion. This competition is evaluated in Fig.~\ref{fig:deterministicD} by considering the deterministic evolution (Eq.~\eqref{eq:FullSDE} without noise term) of an initial Gaussian net-baryon density profile. Using a Gaussian fit, we evaluate the time-evolution of the standard deviation of the density, $y_D(\tau)$, translate it into Cartesian coordinates as $z_D(\tau) = \tau \sinh(y_D(\tau))$ and compare to the increasing size of the expanding system $z_{max}(\tau)$ via the quantity $\widetilde{z}(\tau) = \frac{z_{max}(\tau) - z_D(\tau)}{z_{max}(\tau)}$. \\
\indent If $\widetilde{z}$ is close to 1, the Gaussian initial distribution barely evolves during the expansion and then, in a stochastic context, fluctuations will be correlated over a very small distance and remain out-of-equilibrium. Conversely, if $\widetilde{z} < 0$, fluctuations will be correlated over distances larger than the total system size and then will be closer to equilibrium. Following this observation, three situations emerge. First, for $D \leq 0.5$~fm (blue curves), $\widetilde{z}$ is close to $1$ in the scaling region which implies that the fluctuations are not in equilibrium when passing near the critical point and thus remain roughly unaffected by the criticality. The expansion dominates the diffusion. This small impact of the criticality on the fluctuations will nonetheless be long-lived as can be seen by the large size of the plateau near the proper time $\tau_c - \tau_0 = 2.3$~fm/c where the critical temperature is reached (vertical dashed line). This plateau is due to the critical slowing down, i.e.~the slowdown of the diffusion near the critical point, which here corresponds to an increase of the derivative of $\widetilde{z}$. \\
\indent Second, for $D \geq 0.7$~fm (red curves), $\widetilde{z}(\tau)$ becomes negative before $\tau_c$ is reached. Then, the net-baryon density fluctuations are close to equilibrium in the critical region and are largely impacted by the criticality. The diffusion wins over the expansion. However, as can be seen for the curve $D=0.7$~fm for example, the critical slowing down plateau is now very small meaning that the large critical enhancement of the fluctuations will be very short-lived.\\
\indent Third, for intermediate states $0.5$~fm $ < D < 0.7$~fm (green curves), the fluctuations are close enough to equilibrium to be impacted by the criticality and  the critical signal survives a reasonable amount of time after being printed into the system due to the smaller diffusion. 

\section{Impact on heavy-ion collision phenomenology}
\label{sec:IV}

The diffusive properties of the expanding medium are of crucial importance for critical point studies. More precisely, the evolution of the diffusion length sets the compromise between the amplitude of the critical signal and its longevity. Its impact on heavy-ion collision phenomenology is studied in this work using numerical simulations of Eq.~\eqref{eq:FullSDE}. The evolution of the temperature as a function of proper time in the collision process is modeled in line with a Bjorken expansion as 
\begin{equation}
 T(\tau) = T_i \Big(\frac{\tau_0}{\tau}\Big) \,. 
 \label{eq:Hubble}
\end{equation}

To avoid any additional out-of-equilibrium features, we build up the initial conditions by letting the fluctuations of the net-baryon density evolve at constant temperature $T = 0.5$ GeV for a large number of time steps. When equilibrium is reached, the expansion starts and lasts for a total duration of $5$ fm/c. At a given freeze-out temperature, averages over a rapidity window $\Delta y$ of the net-baryon density fluctuations $\tilde{n}_{\Delta y, i}$ are performed for a large amount of noise configurations $i$. From this distribution the central moments are calculated as 
\begin{equation}
 m_{n, \Delta y}(\tau) = \sum_{i = 0}^{N_{\text{sim}}-1} \big(\tilde{n}_{\Delta y,i}(\tau) - \bar{n}_B \big)^n dy \,,
 \label{eq:mom}
\end{equation}
where $dy$ is the size of the cell used for numerical calculations, $\bar{n}_B$ is the mean value of the distribution of the averages over the different noise configurations and $N_{\text{sim}}$ is the number of simulated noise configurations. Eventually, the second- and fourth-order integrated cumulants $\kappa_{n, \Delta y}$ are obtained from the relation between the central moments and the cumulants for $n = 2, 4$ as
\begin{equation}
 \kappa_{2, \Delta y}(\tau) = m_{2, \Delta y} \,, \quad 
 \kappa_{4, \Delta y}(\tau) = m_{4, \Delta y} - 3 m_{2, \Delta y} \,.
 \label{eq:cumulantrelation}
\end{equation}

\begin{figure}[t!]
    \centering
    \includegraphics[scale=0.22]{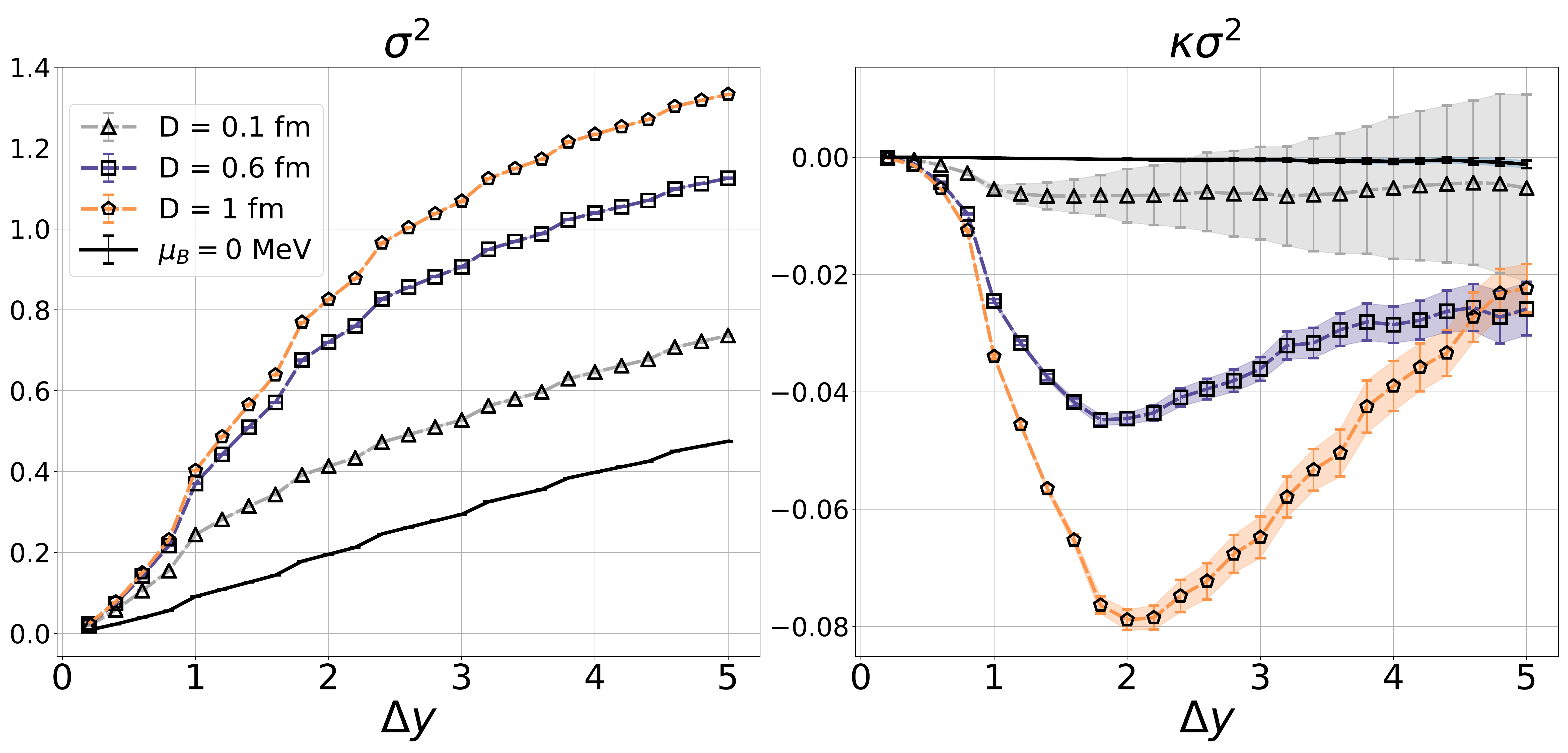}
        \caption{The second- (left panel) and fourth-order (right panel) cumulants as a function of the rapidity window $\Delta y$ at a given freeze-out temperature $T_f = 145$~MeV for trajectories of constant $\mu_B=350$~MeV close to the critical point for the three different situations for the diffusion length discussed in Sec.~\ref{sec:III}. For reference, non-critical cumulants in equilibrium at constant $\mu_B=0$~MeV are shown as black solid curves.}
    \label{fig:varkurt}
\end{figure}

\begin{figure}[t!]
    \centering
    \includegraphics[width=0.45\textwidth,clip]{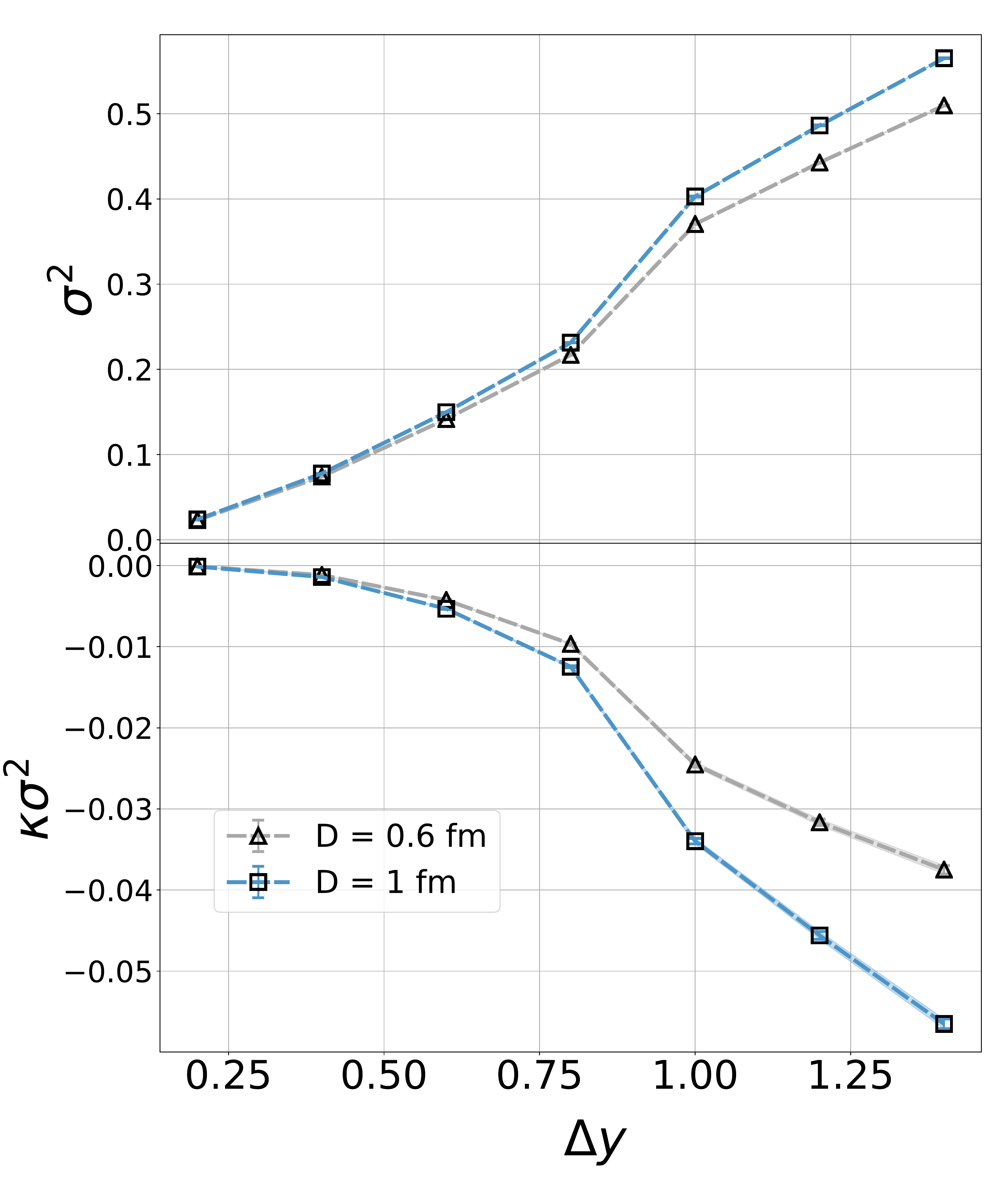}
    \hspace{0.03\textwidth}
    \includegraphics[width=0.45\textwidth,clip]{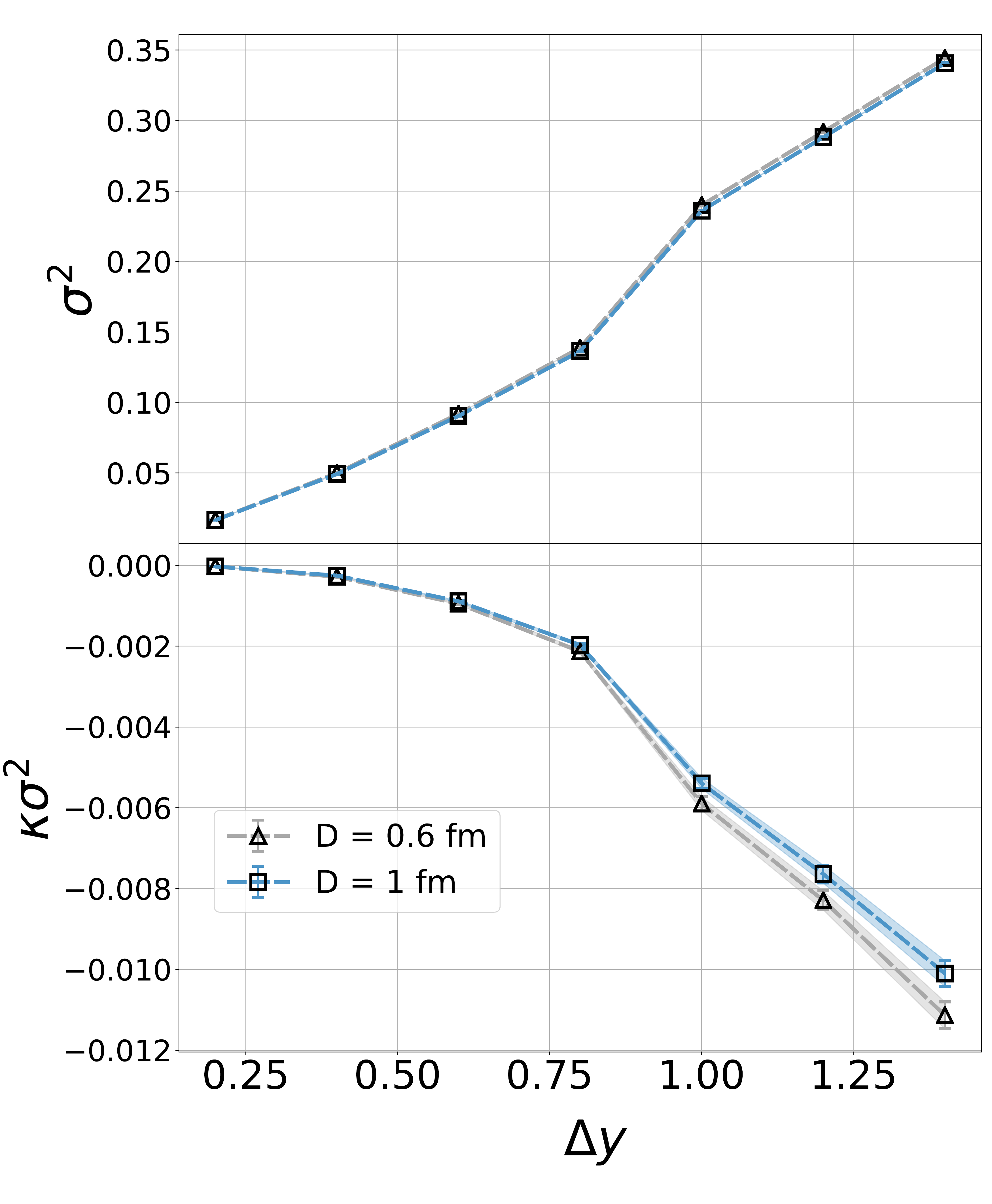}
    \caption{The second- (top) and fourth-order (bottom) cumulants as a function of the rapidity window $\Delta y$ for $D = 0.6, 1$~fm at constant $\mu_B = 350$~MeV for a freeze-out temperature of $T_f = 145$~MeV (left panel) and $T_f = 130$~MeV (right panel).}
    \label{fig:results}
\end{figure}

In Fig.~\ref{fig:varkurt}, we present these fluctuation observables, which are relevant probes of criticality in the QCD phase diagram in line with experimental measurements, at a freeze-out temperature of $T_f = 145$~MeV for the three different regimes discussed in the previous section.
We clearly observe an enhancement of the maximum/minimum of the fluctuations with increasing $D$.
However, this straightforward interpretation is not generally true for smaller rapidity windows and smaller freeze-out temperatures as presented in Fig.~\ref{fig:results}. In fact, going from $T_f = 145$~MeV (left panel) to $T_f = 130$~MeV (right panel) changes the outcome: a smaller $D$ may induce an equivalent or even larger amplitude of the critical signal at $T_f$. Indeed, a smaller diffusion length implies a smaller amplitude but also a longer survival of the signal. 
Thus, having an uncertainty in the value of the diffusion length in addition to an uncertainty in the freeze-out conditions may lead to a misinterpretation of experimental results as either originating from a large but short-lived critical signal or from a small but long-lived one. 
 
As a consequence, a lack of knowledge on the value of the diffusion length $D$ results in a broader range of possible locations for the QCD critical point.

\section{Discussion}

We have seen that the dynamical interplay between the diffusion and the expansion of the medium created in heavy-ion collisions has a strong impact on the net-baryon density fluctuations at freeze-out. In particular, a large diffusion length leads to an enhancement but a small lifetime of the critical signal while for smaller diffusion lengths the opposite is the case. Numerical simulations have shown that a change in the diffusive properties as well as a variation of the freeze-out conditions may lead to a different outcome. This implies that the interpretation of experimental data is very sensitive to the diffusion length. A realistic calculation of the diffusive properties was recently performed in~\cite{Fotakis:2019nbq} and the use of these results for the study of the fluctuation dynamics is currently under investigation.


\end{document}